\documentclass[12pt]{article}
\usepackage{amsfonts,amssymb,amsmath,graphicx,color,epsfig}

\newcommand{\be}{\begin{equation}}
\newcommand{\ee}{\end{equation}}
\newcommand{\bea}{\begin{eqnarray}}
\newcommand{\eea}{\end{eqnarray}}
\newcommand{\beas}{\begin{eqnarray*}}
\newcommand{\eeas}{\end{eqnarray*}}

\newcommand{\phinotg} [1]{{\phi_{0#1}^{\it global}}}

\newcommand{\hw}{\hat{\omega}}
\newcommand{\hk}{\hat{k}}
\newcommand{\htt}{\hat{t}}
\newcommand{\hp}{\hat{\phi}}
\newcommand{\hx}{\hat{x}}
\newcommand{\hy}{\hat{y}}

\begin{document}

\baselineskip 14 pt
\parskip 12 pt

\begin{titlepage}
\begin{flushright}
{\small CU-TP-1149} \\
{\small hep-th/0606141}
\end{flushright}

\begin{center}

\vspace{2mm}

{\Large \bf Holographic representation of local bulk operators}

\vspace{3mm}

Alex Hamilton${}^1$, Daniel Kabat${}^1$, Gilad Lifschytz${}^2$, \\
David A.\ Lowe${}^3$ \\

\vspace{2mm}

${}^1${\small \sl Department of Physics} \\
{\small \sl Columbia University, New York, NY 10027 USA} \\
{\small \tt hamilton@phys.columbia.edu, kabat@phys.columbia.edu}
\vspace{1mm}

${}^2${\small \sl Department of Mathematics and Physics and CCMSC} \\
{\small \sl University of Haifa at Oranim, Tivon 36006 ISRAEL} \\
{\small \tt giladl@research.haifa.ac.il}
\vspace{1mm}

${}^3${\small \sl Department of Physics} \\
{\small \sl Brown University, Providence, RI 02912 USA} \\
{\small \tt lowe@brown.edu}

\end{center}

\vskip 0.3 cm

\noindent
The Lorentzian AdS/CFT correspondence implies a map between local
operators in supergravity and non-local operators in the CFT.  By
explicit computation we construct CFT operators which are dual to
local bulk fields in the semiclassical limit.  The computation is done
for general dimension in global, Poincar\'e and Rindler coordinates.
We find that the CFT operators can be taken to have compact support in
a region of the complexified boundary whose size is set by the bulk
radial position.  We show that at finite $N$ the number of independent
commuting operators localized within a bulk volume saturates the
holographic bound.

\end{titlepage}

\section{Introduction}

The anti-de Sitter/conformal field theory (AdS/CFT) correspondence
\cite{Maldacena:1997re,Gubser:1998bc,Witten:1998qj,Aharony:1999ti}, in
its Lorentzian version
\cite{Balasubramanian:1998sn,Banks:1998dd,Balasubramanian:1998de},
states that any bulk excitation is encoded on the boundary by some CFT
operator or state. In the semiclassical limit of large $N$ and large
't Hooft coupling we expect to have free local fields in the
bulk. These bulk fields should be encoded in the CFT.  To see how this
works consider a bulk field with normalizable fall-off near the
boundary of AdS.
\[
\phi(z,x) \sim z^\Delta \phi_0(x)
\]
Here $z$ is a radial coordinate which vanishes at the boundary.  The
bulk supergravity field can be expressed in terms of the boundary
field $\phi_0$ via a kernel $K$.
\[
\phi(z,x) = \int dx' \, K(x' \vert z,x) \phi_0(x')
\]
We will refer to $K$ as a smearing function.  $\phi_0(x)$ corresponds
to a local operator ${\cal O}(x)$ in the CFT \cite{Klebanov:1999tb}.
\[
\phi_0(x) \leftrightarrow {\cal O}(x)
\]
Thus the AdS/CFT correspondence implies that local bulk fields are
dual to non-local boundary operators
\cite{Banks:1998dd,Balasubramanian:1999ri,Bena}.\footnote{For a
different approach see \cite{Rehren:1999jn,Rehren:2000tp}.}
\begin{equation}
\phi(z,x) \leftrightarrow \int dx' \, K(x' \vert z,x){\cal O}(x')
\label{basic}
\end{equation} 
Bulk-to-bulk correlation functions, for example, are equal to
correlation functions of the corresponding non-local operators in the
CFT.
\[
\langle \phi(z_1,x_1) \phi(z_2,x_2) \rangle_{SUGRA} = \int dx_1' dx_2' \,
K(x_1' \vert z_1,x_1) K(x_2' \vert z_2,x_2) \langle {\cal O}(x_1')
{\cal O}(x_2') \rangle_{CFT}
\]

Smearing functions are central to understanding Lorentzian AdS/CFT:
they define the map by which, in the semiclassical limit, local bulk
excitations are encoded on the boundary.  The semiclassical limit
tightly constrains behavior at finite $N$.  For example, as we will
see, smearing functions can be used to count the number of independent
commuting operators inside a volume in the bulk, even at finite $N$.
They can also be used to study bulk locality and causality: for
example in \cite{hkll} we used them to understand the causal structure
of a black hole from the boundary point of view.

The main purpose of this paper is to compute smearing functions in
various pure AdS geometries.  This continues the study started in
\cite{hkll}, where a two-dimensional AdS spacetime was considered. In
the present paper we extend the analysis to higher dimensions and
compute smearing functions for global AdS${}_{d+1}$, for the
Poincar\'e patch in $d+1$ dimensions and for AdS${}_3$ in Rindler
coordinates.

It's important to recognize that smearing functions are not
necessarily unique.  In some cases the boundary fields do not involve
a complete set of Fourier modes and we are free to add to the smearing
function terms that integrate to zero against all boundary
fields. This freedom enables us to present the smearing function in
different forms, which is useful depending on which aspect one wishes
to study.

\begin{figure}
\centerline{\includegraphics{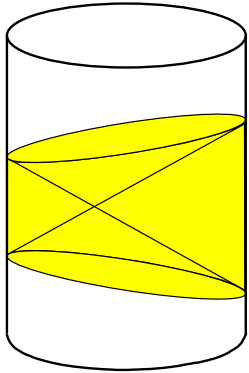}}
\caption{In global coordinates AdS resembles an infinite cylinder.
We've drawn the light cones emanating from a bulk point and
intersecting the boundary.  The CFT operator has support on the
strip indicated in yellow, at spacelike separation from the bulk
point.}
\label{GlobalFig}
\end{figure}

For the impatient reader, let us briefly summarize our main results.
In global coordinates we find that the smearing function can be chosen
to have support on boundary points that are spacelike separated from
the bulk point.  This is illustrated in figure \ref{GlobalFig}.  The
exact form of the smearing function depends on the dimension: for
even-dimensional AdS it's given in (\ref{GlobalEvenSmear}), for
odd-dimensional AdS it's given in (\ref{GlobalOddSmear}).

We also construct smearing functions in the Poincar\'e patch.  For
even-dimensional AdS we find that the smearing function can be taken
to have support at spacelike separation in the Poincar\'e patch: see
(\ref{PoincareEvenSmear}).  For odd-dimensional AdS the smearing
function has support on the entire Poincar\'e boundary: see
(\ref{PoincareOddSmear}).  An alternate form of the AdS${}_3$
Poincar\'e smearing function is given in (\ref{PoincareModeSum}).

In Rindler coordinates we show that to construct a smearing function
we must analytically continue the spatial coordinates of the boundary
theory to imaginary values.  This enables us to find a smearing
function with support on a compact region of the complexified
geometry.  The explicit result for AdS${}_3$ is given in
(\ref{finalanswer}).

It is desirable to work with smearing functions that make the boundary
operators as sharply-localized as possible.  Our strongest results in
this direction are achieved in Rindler coordinates, where we find
operators with compact support on the complexified boundary.  This
enables us to obtain an improved understanding of bulk causality.
Furthermore the statement of scale-radius duality
\cite{Susskind:1998dq,Peet:1998wn} can be made in a sharper way, since
the size of the smeared operator is determined by the radial position
of the bulk point. We would also like to stress that, as in
\cite{hkll}, lightcone singularities of bulk correlators arise from UV
singularities of the boundary theory.  This is true even for bulk
points deep inside AdS.  Thus regions deep inside the bulk cannot be
associated with a boundary theory with a conventional UV cutoff.

An outline of this paper is as follows.  In section \ref{gads} we
compute the smearing function in global coordinates.  In section
\ref{pads} we compute the smearing function in Poincar\'e coordinates.
In section \ref{rads} we compute the smearing function for AdS${}_3$
in Rindler coordinates, and in section \ref{physics} we discuss some
of the implications of our results for bulk locality and holography.
Finally appendix \ref{eggads} presents an alternate derivation of the
global smearing function in even-dimensional AdS, appendix
\ref{covariance} shows that the smearing functions are AdS-covariant,
and appendix \ref{Poincare3} presents an alternate derivation of the
Poincar\'e smearing function in AdS${}_3$.

\section{Global AdS} \label{gads} 

In this section we construct smearing functions in global coordinates.
The construction is based on mode sums.  We treat even-dimensional AdS
in section \ref{egads} and odd-dimensional AdS in section \ref{ogads}.
In appendix \ref{eggads} we present an alternate approach to the
even-AdS global smearing function, where the construction is based on
a Greens function.

\subsection{Preliminaries}

We will describe AdS${}_{D = d+1}$ in global coordinates
\begin{equation}
ds^2 = \frac{R^2} { \cos^2 \rho} \left(-d\tau^2 + d\rho^2 + \sin^2 \! \rho \, d\Omega_{d-1}^2\right)
\end{equation}
where $R$ is the AdS radius, $- \infty < \tau < \infty$, $0 \leq \rho < \pi/2$, and $d\Omega_{d-1}^2$ is the
metric on a unit $(d-1)$--sphere.  An AdS-invariant distance function is given by
\begin{equation}
\label{DistanceFtn}
\sigma(x \vert x') = \frac{\cos(\tau - \tau') - \sin \rho \sin \rho' \cos(\Omega - \Omega') }{ \cos \rho \cos \rho'}
\end{equation}
where $\Omega - \Omega'$ is the angular separation on the sphere.  For $- \pi < \tau - \tau' < \pi$ points with $\sigma > 1$
are spacelike separated, while points with $\sigma = 1$ are lightlike separated and points with $\sigma < 1$ are
timelike separated (although they can only be connected by a timelike geodesic if $-1 < \sigma < 1$).  Solutions to the wave
equation $(\Box - m^2)\phi = 0$ can be expanded in normalizeable modes
\begin{equation}
\label{GlobalModes}
\phi(\tau,\rho,\Omega) = \sum_{n = 0}^\infty \sum_{l,m} a_{nlm} e^{-i(2n + l + \Delta)\tau}
\cos^\Delta \! \rho \sin^l \! \rho P_n^{\left(\Delta - \frac{d}{2},\,l + \frac{d}{2} - 1\right)}
(-\cos 2 \rho) Y_{lm}(\Omega)
+ {\rm c.c.}
\end{equation}
where $P_n^{(\alpha,\beta)}$ is a Jacobi polynomial, $Y_{lm}$ is a spherical harmonic, and the conformal dimension of the
corresponding operator is
$\Delta = \frac{d}{2} + \sqrt{{d^2 \over 4} + m^2 R^2}$.  In global coordinates we define the boundary value of the field
\begin{equation}
\label{gbmode}
\phinotg{}(\tau,\Omega) = \lim_{\rho \rightarrow \pi/2} {\phi(\tau,\rho,\Omega) \over \cos^\Delta \rho}\,.
\end{equation}

\subsection{Even AdS: mode sum approach} \label{egads}

We assume that $D$ is even and construct a smearing function starting from the mode expansion (\ref{GlobalModes}).
We first work at the center of AdS (meaning $\rho = 0$), where only the $s$-wave contributes, and later extend
our results to arbitrary bulk points.  At the center
\begin{equation}
\label{GlobalSwave}
\phi(\tau,\rho = 0,\Omega) = \sum_{n = 0}^\infty a_n e^{-i(2n+\Delta)\tau} P_n^{\left(\Delta - {d \over 2},\,
{d \over 2} - 1\right)}(-1) + {\rm c.c.}
\end{equation}
We can split the corresponding $s$-wave part of the boundary field into its positive and negative frequency components,
\begin{eqnarray}
&&\phinotg{}(\tau) = \phinotg{+}(\tau) + \phinotg{-}(\tau) \\
\label{GlobalSplus}
&&\phinotg{+} = \sum_{n = 0}^\infty a_n e^{-i(2n+\Delta)\tau} P_n^{\left(\Delta - {d \over 2},\,{d \over 2} - 1\right)}(1) \\
\label{GlobalSminus}
&&\phinotg{-} = \sum_{n = 0}^\infty a_n^* e^{i(2n+\Delta)\tau} P_n^{\left(\Delta - {d \over 2},\,{d \over 2} - 1\right)}(1)~.
\end{eqnarray}
Note that
\begin{equation}
a_n = {1 \over \pi \, {\rm vol}(S^{d-1}) P_n^{\left(\Delta - {d \over 2},\,{d \over 2} - 1\right)}(1)} \,
\int_{-\pi/2}^{\pi/2} d\tau \int d\Omega \sqrt{g_\Omega} \, e^{i(2n+\Delta)\tau} \phinotg{+}(\tau) ~.
\end{equation}
Plugging this back into the bulk mode expansion (\ref{GlobalSwave}) we can write the field at the origin of AdS
(meaning the point $\tau' = \rho' = 0$) as
\begin{equation}
\label{OddOrigin}
\phi \vert_{\rm origin} = \int_{-\pi/2}^{\pi/2} d\tau \int d\Omega \sqrt{g_\Omega} \, K_+(\tau,\Omega \vert \tau',\rho',\Omega')
\phinotg{+}(\tau,\Omega) + {\rm c.c.}
\end{equation}
where
\begin{equation}
K_+ = {1 \over \pi {\rm vol}(S^{d-1})} \, \sum_{n = 0}^\infty e^{i (2n+\Delta) \tau}  {P_n^{\left(\Delta - {d \over 2},\,
{d \over 2} - 1\right)}(-1) \over P_n^{\left(\Delta - {d \over 2},\, {d \over 2} - 1\right)}(1)}\,.
\end{equation}
The sum can be evaluated as
\begin{eqnarray}
\nonumber
K_+ & = & {\Gamma(\Delta - {d \over 2} + 1) \over \pi {\rm vol}(S^{d-1}) \Gamma(d/2)} \, e^{i \Delta \tau} \,
\sum_{n = 0}^\infty {\Gamma(n + {d \over 2}) \over \Gamma(n + \Delta - {d \over 2} + 1)} (-e^{i 2 \tau})^n \\
\label{CenterModeSum}
& = & {1 \over \pi {\rm vol}(S^{d-1})} \, e^{i \Delta \tau} F\Bigl(1, {d \over 2},\Delta - {d \over 2} + 1,-e^{i 2 \tau}\Bigr)
\end{eqnarray}
where strictly speaking to make the sum convergent we should have replaced $\tau \rightarrow \tau + i \epsilon$.

In terms of $z = e^{i 2 \tau}$ we have $K_+ = {1 \over \pi {\rm vol}(S^{d-1})} \, z^{\Delta / 2}
F(1, {d \over 2},\Delta - {d \over 2} + 1,-z)$.  At this point it's useful to make a $z \rightarrow 1/z$
transformation of the hypergeometric function.  This gives
\begin{eqnarray}
\nonumber
&& K_+ = {z^{\Delta/2} \over \pi {\rm vol}(S^{d-1})} \, \Biggl\lbrace
{\Gamma(\Delta - {d \over 2} + 1) \Gamma({d \over 2} - 1) \over \Gamma(d/2) \Gamma(\Delta - d/2)}
z^{-1} F\left(1,\,1+{d \over 2} - \Delta,\, 2 - {d \over 2},\, -{1 \over z}\right) \\
\label{XformedKplus}
& & \qquad\qquad +
{\Gamma(\Delta - {d \over 2} + 1) \Gamma(1 - {d \over 2}) \over \Gamma(1) \Gamma(\Delta - d + 1)}
z^{-d/2} F\left(d - \Delta,\, {d \over 2},\, {d \over 2},\, - {1 \over z}\right)
\Biggr\rbrace~.
\end{eqnarray}
It's important to note that smearing functions aren't unique, since we could replace
\begin{equation}
K_+ \rightarrow K_+ + z^{\Delta/2} \sum_{n = 1}^\infty c_n z^{-n}
\end{equation}
for any set of $c_n$: the extra terms involve Fourier components which
are absent from the mode expansion (\ref{GlobalSplus}), so they drop
out when integrated against $\phinotg{+}$.  This freedom can be used
to eliminate the first line in (\ref{XformedKplus}), as can be seen by
expanding the hypergeometric function there in powers of
$1/z$.\footnote{We're cheating a bit here, given our $\tau \rightarrow
\tau + i \epsilon$ prescription, since the hypergeometric series
only converges inside the unit disc.  This can be taken into account
by slightly deforming the $z$ contour of integration.}  Then using
$F(\alpha,\beta,\beta,x) = (1 - x)^{-\alpha}$ in the second line of
(\ref{XformedKplus}) we are left with
\begin{equation}
K_+ = {\Gamma(\Delta - {d \over 2} + 1) \Gamma(1 - {d \over 2}) \over \pi {\rm vol}(S^{d-1}) \Gamma(\Delta - d + 1)}
\left(\sqrt{z} + {1 \over \sqrt{z}}\right)^{\Delta - d}\,.
\end{equation}
Note that $K_+$ is real, so we can set $K = K_+ = K_-$.  The full smearing function for a bulk point at the origin
is then given by
\be
K = {\Gamma(\Delta - {d \over 2} + 1) \Gamma(1 - {d \over 2}) \over \pi {\rm vol}(S^{d-1}) \Gamma(\Delta - d + 1)}
\left(2 \cos \tau \right)^{\Delta - d}\,.
\ee
It's useful to express this in terms of the invariant distance (\ref{DistanceFtn}).
In global coordinates the regulated distance from the origin of AdS to a point on the boundary is
$\lim_{\rho \rightarrow \pi/2} \sigma \cos \rho = \cos \tau$.  In terms of this regulated distance the smearing
function for a bulk point at the origin is
\be
\label{OriginSmearEven}
K = {\Gamma(\Delta - {d \over 2} + 1) \Gamma(1 - {d \over 2}) \over \pi {\rm vol}(S^{d-1}) \Gamma(\Delta - d + 1)}
\lim_{\rho \rightarrow \pi/2} (2 \sigma \cos \rho)^{\Delta - d}\,.
\ee

To extend this to an arbitrary bulk point $P$ we can first use an AdS isometry to move $P$ to the origin, then apply
the smearing function (\ref{OriginSmearEven}) to the transformed boundary data.  Alternatively we can use the original
boundary data but transform the smearing function.  This is straightforward because (\ref{OriginSmearEven}) is
AdS covariant.  Thus for an arbitrary bulk point we have
\begin{equation}
\phi(P) = \int_{-\infty}^{\infty} d\tau \int d\Omega \sqrt{g_\Omega} \, K(\tau,\Omega \vert P) \phinotg{}(\tau,\Omega)
\end{equation}
where
\begin{eqnarray}
\nonumber
&&K(\tau,\Omega \vert P) = c_{d \Delta} \lim_{\rho \rightarrow \pi/2} (\sigma(x \vert P) \cos \rho)^{\Delta - d}
\theta({\rm spacelike}) \\
\label{GlobalEvenSmear}
&& c_{d \Delta} = {(-1)^{(d - 1)/2} 2^{\Delta - D} \Gamma(\Delta - {d \over 2} + 1)
\over \pi^{d/2} \Gamma(\Delta - d + 1)}\,.
\end{eqnarray}
In appendix \ref{eggads} we reproduce this result by constructing a Greens function for the bulk wave equation.

\subsection{Odd AdS: mode sum approach} \label{ogads}

We now assume that $D$ is odd.  As in the previous subsection we first
work at the origin of AdS (meaning $\tau = \rho = 0$), where only the
$s$-wave contributes, and later extend our results to arbitrary bulk
points.

The result (\ref{CenterModeSum}) holds in any number of dimensions, so in terms of $z = e^{i 2 \tau}$ we have the
positive-frequency part of the global smearing function for a bulk point at the origin of AdS
\[
K_+ = {1 \over \pi {\rm vol}(S^{d-1})} \, z^{\Delta / 2} F(1, {d \over 2},\Delta - {d \over 2} + 1,-z)\,.
\]
At this point it's useful to make a $z \rightarrow 1/z$ transformation of the hypergeometric function.  Noting that
${d \over 2}$ is an integer, the relevant formula can be found in \cite{Bateman}, p.~109 equation (7).  Again it's
important to note that the smearing functions aren't unique, since we could replace
\begin{equation}
K_+ \rightarrow K_+ + z^{\Delta/2} \sum_{n = 1}^\infty c_n z^{-n}
\end{equation}
for any $c_n$ since the extra terms drop out when integrated against $\phinotg{+}$.  Making the $z \rightarrow 1/z$
transformation and dropping terms that don't contribute we are left with
\begin{equation}
K_+ = - {z^{(\Delta - d)/2} \log z \over \pi {\rm vol}(S^{d-1}) \Gamma(d/2) \Gamma(d/2 - \Delta)} \,
\sum_{n = 0}^\infty {1 \over n!} \Gamma(n + d - \Delta) (-z)^{-n}\,.
\end{equation}
With some transformations of the gamma function, the binomial series can be rewritten as
\begin{equation}
(1 + x)^\alpha = - {1 \over \pi} \sin (\pi \alpha) \, \Gamma(\alpha + 1) \sum_{n = 0}^\infty {1 \over n!} \Gamma(n - \alpha)
(-x)^n\,.
\end{equation}
So in fact
\begin{equation}
\label{KplusOrigin}
K_+ = {(-1)^{(d - 2)/2} \Gamma(\Delta - {d \over 2} + 1) \over 2 \pi^{1 + {d \over 2}} \Gamma(\Delta - d + 1)}
\lim_{\rho \rightarrow \pi/2} (2 \sigma \cos \rho)^{\Delta - d} \log z
\end{equation}
where we introduced the invariant distance (\ref{DistanceFtn}) from the origin of AdS to a point on
the boundary via
\begin{equation}
\lim_{\rho \rightarrow \pi/2} 2 \sigma \cos \rho = \sqrt{z} + {1 \over \sqrt{z}}\,.
\end{equation}
Using (\ref{KplusOrigin}) in (\ref{OddOrigin}), we can express the value of the field at the origin of AdS as
\begin{equation}
\phi \vert_{\rm origin} = A \int_{-\pi/2}^{\pi/2} d\tau \int d\Omega \sqrt{g_\Omega} \lim_{\rho \rightarrow \pi/2} \,
(2 \sigma(x \vert x') \cos \rho)^{\Delta - d} \log z \left(\phinotg{+}(\tau,\Omega) - \phinotg{-}(\tau,\Omega)\right)
\end{equation}
where $A = {(-1)^{(d - 2)/2} \Gamma(\Delta - {d \over 2} + 1) \over 2 \pi^{1 + {d \over 2}} \Gamma(\Delta - d + 1)}$.
This is progress, but we'd like to express $\phi$ in terms of the local combination $\phinotg{} = \phinotg{+} + \phinotg{-}$.
To do this it's useful to note that for even $d$
\begin{equation}
\lim_{\rho \rightarrow \pi/2} \, (2 \sigma \cos \rho)^{\Delta - d} = z^{\Delta / 2} z^{-d/2} \left(1 + {1 \over z}\right)^{\Delta - d}
\end{equation}
has an expansion in inverse powers of $z$,
\begin{equation}
\label{EvenIdentity}
\lim_{\rho \rightarrow \pi/2} \, (2 \sigma \cos \rho)^{\Delta - d} = z^{\Delta / 2} \sum_{n = 1}^\infty c_n z^{-n}\,.
\end{equation}
A function of this form vanishes when integrated against $\phinotg{+}$.  Likewise, by expanding in positive
powers of $z$, it vanishes when integrated against $\phinotg{-}$.  So we have the identity\footnote{This identity
shows that a global smearing function of the form one might have naively expected, namely $K \sim
(\sigma\cos\rho)^{\Delta - d}$, cannot be correct in odd-dimensional AdS.}
\begin{equation} \label{zisz}
\int_{-\pi/2}^{\pi/2} d\tau \int d\Omega \sqrt{g_\Omega} \lim_{\rho \rightarrow \pi/2} \, (\sigma \cos \rho)^{\Delta - d}
\left(\phinotg{+}(\tau,\Omega) + \phinotg{-}(\tau,\Omega)\right) = 0\,.
\end{equation}
Differentiating this identity with respect to $\Delta$, including the factors of $z^{\pm \Delta/2}$ hidden in the
mode expansion of $\phinotg{\mp}$, we obtain\footnote{We differentiate with respect to $\Delta$ holding the quantities
$a_n P_n^{(\Delta - {d \over 2},{d \over 2} - 1)}(1)$, $a_n^* P_n^{(\Delta - {d \over 2},{d \over 2} - 1)}(1)$ which
appear in the mode expansions fixed.  See (\ref{GlobalSplus}), (\ref{GlobalSminus}).}
\begin{eqnarray}
\nonumber
&& \int_{-\pi/2}^{\pi/2} d\tau \int d\Omega \sqrt{g_\Omega} \lim_{\rho \rightarrow \pi/2} \,
(\sigma \cos \rho)^{\Delta - d} \log z (\phinotg{+} - \phinotg{-}) \\
&& = 2 \int_{-\pi/2}^{\pi/2} d\tau \int d\Omega \sqrt{g_\Omega} \lim_{\rho \rightarrow \pi/2} \,
(\sigma \cos \rho)^{\Delta - d} \log (\sigma \cos \rho) \, \phinotg{}\,.
\end{eqnarray}
This lets us express the value of the field at the origin of AdS in terms of an integral over points on the
boundary that are spacelike separated from the origin:
\begin{equation}
\label{origin}
\phi \vert_{\rm origin} = 2A \int_{-\pi/2}^{\pi/2} d\tau \int d\Omega \sqrt{g_\Omega} \lim_{\rho \rightarrow \pi/2} \,
(2 \sigma \cos \rho)^{\Delta - d} \log (\sigma \cos \rho) \, \phinotg{}\,.
\end{equation}

Finally we'd like to extend these results to an arbitrary bulk point.  We claim that
\begin{equation} \label{ogsmear}
\phi(x') = \int_{-\infty}^{\infty} d\tau \int d\Omega \sqrt{g_\Omega} \, K(\tau,\Omega \vert x') \phinotg{}(\tau,\Omega)
\end{equation}
where
\begin{eqnarray}
\nonumber
K(\tau,\Omega \vert x') &=& a_{d \Delta} \lim_{\rho \rightarrow \pi/2} (\sigma(x \vert x') \cos \rho)^{\Delta - d}
\log \big(\sigma(x \vert x') \cos \rho\big) \theta({\rm spacelike}) \\
\label{GlobalOddSmear}
 a_{d \Delta} &=& {(-1)^{(d - 2)/2} 2^{\Delta - d} \Gamma(\Delta - {d \over 2} + 1)
\over \pi^{1 + {d \over 2}} \Gamma(\Delta - d + 1)}\,.
\end{eqnarray}
The argument is as follows.  To compute the field at $x'$ one can
first use an AdS isometry to move $x'$ to the origin, then use the
smearing function (\ref{origin}) to compute $\phi$ at the origin in
terms of the transformed boundary data.  Equivalently, one can use the
original boundary data but transform the smearing function.  This is
easy to do because, as we show in appendix \ref{covariance},
(\ref{ogsmear}) is secretly AdS covariant.

\section{Poincar\'e smearing} \label{pads}

In Poincar\'e coordinates the AdS metric is
\[
ds^2 = {R^2 \over Z^2} \left(-dT^2 + \vert d\vec{X} \vert^2 + dZ^2\right)
\]
where $R$ is the AdS radius and $0 < Z < \infty$.  These coordinates
cover a wedge-shaped region of global AdS.  In section \ref{epads} we
work in even-dimensional AdS and construct a spacelike smearing
function with support in the Poincar\'e patch, starting from our
global result (\ref{GlobalEvenSmear}).  In section
\ref{sect:OddAdSPoincare} we follow the same procedure in
odd-dimensional AdS, and find that for generic $\Delta$ it leads to a
smearing function with support on the entire Poincar\'e boundary.  In
appendix \ref{Poincare3} we present an alternate form of the smearing
function for AdS${}_3$, based on mode sums in the Poincar\'e patch.

\subsection{Even AdS} \label{epads}

In even AdS we can construct a spacelike smearing function with
support in a Poincar\'e patch, starting from the global spacelike
smearing function (\ref{GlobalEvenSmear}).

In global coordinates the antipodal map acts via \footnote{It's
defined on the AdS hyperboloid, so it's ambiguous whether $\tau$ is
increased or decreased by $\pi$.  We'll need both options below.}
\[
A \, : \, \tau \rightarrow \tau \pm \pi\, , \qquad \rho ~ {\rm invariant} \, ,\qquad \Omega \rightarrow \Omega_A
\]
where $\Omega_A$ is the antipodal point on the sphere.  The positive-frequency part of a bulk field transforms by
\[
\phi_+(Ax) = e^{\mp i \pi \Delta} \phi_+(x)
\]
under the antipodal map.

Given a bulk point $P$ contained inside some Poincar\'e patch, the
global smearing function consists of three regions on the boundary.
Region I is located to the past of the Poincar\'e patch, region II is
contained within the Poincar\'e patch, and region III is to the future
of the Poincar\'e patch.  By applying a $\tau \rightarrow \tau + \pi$
antipodal map to region I, and a $\tau \rightarrow \tau - \pi$
antipodal map to region III, everything gets mapped inside the
Poincare patch.  Thus we can re-write the global smearing function as
\begin{eqnarray*}
\phi(P) & = & \int d\tau d\Omega \, K_{\rm global}(\tau,\Omega \vert P) (\phi_{0+}^{\rm global} + \phi_{0-}^{\rm global}) \\
& = & \int_{{\scriptstyle \rm Poincare} \atop {\scriptstyle \rm patch}} \hspace{-1cm} d\tau d\Omega \,
c_{d\Delta} \vert \sigma \cos \rho \vert^{\Delta - d}
\left\lbrace \begin{array}{ll}
e^{i \pi \Delta}  \phi_{0+}^{\rm global} & \quad \hbox{\rm in image of region I} \\
                  \phi_{0+}^{\rm global} & \quad \hbox{\rm in region II} \\
e^{-i \pi \Delta} \phi_{0+}^{\rm global} & \quad \hbox{\rm in image of region III}
\end{array}\right\rbrace + {\rm c.c.}
\end{eqnarray*}
where $c_{d\Delta}$ is the constant given in (\ref{GlobalEvenSmear}).
We've used the fact that the integration measure is invariant under
the antipodal map while $\sigma(x \vert x') = - \sigma(Ax \vert x')$.
By regarding the phases as part of the smearing function rather than
as part of the boundary field we have
\[
\phi(P) = \int_{{\scriptstyle \rm Poincare} \atop {\scriptstyle \rm patch}} \hspace{-1cm} d\tau d\Omega \, c_{d\Delta}
\left\lbrace\begin{array}{c}
e^{i \pi \Delta} \\
1 \\
e^{-i \pi \Delta}
\end{array}\right\rbrace
\vert \sigma \cos \rho \vert^{\Delta - d}
\phi_{0+}^{\rm global} + {\rm c.c.}
\]
Putting in the Jacobians to convert from global to Poincar\'e coordinates, namely
\[
{d\tau d\Omega \over \cos^d \rho} = {dT d^{d-1} X \over Z^d}
\]
and
\[
\cos^\Delta \rho \, \phi_0^{\rm global} = Z^\Delta \phi_0^{\rm Poincare}\,,
\]
this becomes
\begin{equation}
\label{Poincare1}
\phi(P) = \int dT d^{d-1}X \, c_{d\Delta}
\left\lbrace\begin{array}{c}
e^{i \pi \Delta} \\
1 \\
e^{-i \pi \Delta}
\end{array}\right\rbrace
\vert \sigma Z \vert^{\Delta - d}
\phi_{0+}^{\rm Poincare} + {\rm c.c.}
\end{equation}
Now consider the function
\[
f(T,X \vert P) = \lim_{Z \rightarrow 0} \bigl(\sigma(T,Z,X \vert P) Z\bigr)^{\Delta - d}
\]
defined with the prescription $T \rightarrow T - i \epsilon$.  That is,
\begin{eqnarray*}
f(T) & = & \left({1 \over 2 Z'} \bigg(Z'^2 + \vert X - X' \vert^2 - (T - T' - i \epsilon)^2 \bigg)\right)^{\Delta - d} \\
& = & \left\lbrace\begin{array}{ll}
- e^{i \pi \Delta} \vert \sigma Z \vert^{\Delta - d} & \quad \hbox{\rm in image of region I} \\
                   \vert \sigma Z \vert^{\Delta - d} & \quad \hbox{\rm in region II} \\
- e^{-i \pi \Delta} \vert \sigma Z \vert^{\Delta - d} & \quad \hbox{\rm in image of region III}
\end{array}\right.
\end{eqnarray*}
where we have used the fact that $d$ is odd.  Since $f$ is analytic in
the lower half complex $T$ plane its Fourier transform $f(\omega) =
\int dT \, e^{i \omega T} f(T)$ vanishes for $\omega < 0$, and hence $\int
dT \, f(T) \phi_{0+}^{\rm Poincare} = 0$.  We are therefore free to
modify the Poincare smearing function by replacing $K_{+} \rightarrow
K_{+} + c_{d\Delta} f$ in (\ref{Poincare1}).  This exactly cancels the
smearing function at timelike separation, while giving a factor of two
at spacelike separation, resulting in smearing function which is real.
Thus in the end we obtain the Poincar\'e smearing function
\begin{eqnarray}
\label{PoincareEvenSmear}
&&\phi(P) = \int dT d^{d-1}X \, K_{\rm Poincare}(T,X \vert P) \phi_0^{\rm Poincare}(T,X) \\
\nonumber
&&K_{\rm Poincare} = 2 c_{d\Delta} \lim_{Z \rightarrow 0} \left(\sigma(T,Z,X \vert P) Z\right)^{\Delta - d}
\theta({\rm spacelike})\,.
\end{eqnarray}

Note that this smearing function grows at large spacelike separation.
However a boundary field which is globally well-defined must fall off
at large spacelike separation, $\phi_0^{\rm Poincare} \sim
\sigma^{-\Delta}$ as $X \rightarrow \infty$. So the convolution $\int
K \phi_0$ is well-defined.

\subsection{Odd AdS}
\label{sect:OddAdSPoincare}

We now construct a Poincar\'e smearing function in odd-dimensional
AdS, following the same procedure as in the last section: we start
with the global result (\ref{GlobalOddSmear}) and use the antipodal
map to transform it into a Poincar\'e patch.

Applying the same logic as in the last section, we have
\begin{eqnarray*}
& & \phi(P) = \int d\tau d\Omega \, K_{\rm global}(\tau,\Omega \vert P) (\phi_{0+}^{\rm global} + \phi_{0-}^{\rm global}) \\
& & \quad = \int_{{\scriptstyle \rm Poincare} \atop {\scriptstyle \rm patch}} \hspace{-1cm} d\tau d\Omega \, a_{d\Delta}
\vert \sigma \cos \rho \vert^{\Delta - d} \log \vert \sigma \cos \rho \vert
\left\lbrace \begin{array}{ll}
e^{i \pi \Delta}  \phi_{0+}^{\rm global} & \quad \hbox{\rm in image of region I} \\
                  \phi_{0+}^{\rm global} & \quad \hbox{\rm in region II} \\
e^{-i \pi \Delta} \phi_{0+}^{\rm global} & \quad \hbox{\rm in image of region III}
\end{array}\right\rbrace + {\rm c.c.}
\end{eqnarray*}
where $a_{d\Delta}$ is the constant given in (\ref{GlobalOddSmear}).  Again, regarding the phases as part of the
smearing function rather than as part of the boundary field and transforming to Poincar\'e coordinates, we have
\begin{equation}
\phi(P) = \int dT d^{d-1}X \, a_{d\Delta}
\left\lbrace\begin{array}{c}
e^{i \pi \Delta} \\
1 \\
e^{-i \pi \Delta}
\end{array}\right\rbrace
\vert \sigma Z \vert^{\Delta - d} \log \vert \sigma Z \vert
\phi_{0+}^{\rm Poincare} + {\rm c.c.}
\end{equation}
This relies on the fact that, as shown in appendix \ref{covariance}, we can replace
$\log \vert \sigma \cos \rho \vert$ with $\log \vert \sigma Z \vert$ in the smearing
function.

The phases in the smearing function can be absorbed into an $i \epsilon$ prescription.  That is, we have
\begin{equation}
\phi(P) = \int dT d^{d-1}X \, K_+ \phi_{0+}^{\rm Poincare} + {\rm c.c.}
\end{equation}
where
\begin{equation}
K_+ = a_{d\Delta} (\sigma Z)^{\Delta - d} \Big\vert_{T \rightarrow T - i \epsilon} \log \vert \sigma Z \vert\,.
\end{equation}
A function analytic in the lower half $T$ plane gives vanishing result when integrated against $\phi_{0+}^{\rm Poincare}$.
So we can even take $K_+$ to be given by the rather peculiar $i \epsilon$ prescription
\begin{equation}
\label{PoincareOddSmear}
K_+ = {1 \over 2} a_{d \Delta} (\sigma Z)^{\Delta - d} \Big\vert_{T \rightarrow T - i \epsilon} \log (\sigma Z)
\Big\vert_{T \rightarrow T + i \epsilon}\,.
\end{equation}
Note that $K_+$ isn't real in general, so we can't take $K_+ = K_- = K$.  Rather one must first decompose $\phi_0^{\rm
Poincare}$ into its positive and negative frequency components before using these results.  Also note that the smearing
function is not restricted to spacelike separation.  It is, however, AdS covariant.

It is not clear to us whether these peculiar features are fundamental to Poincar\'e smearing in odd dimensions, or can be
overcome in some manner.  However we would like to point out one exceptional case: if $\Delta$ is an integer then $K_+$
can be taken to be real and we have
\begin{equation}
\label{IntegerDeltaPoincare}
\phi(P) = \int dT d^{d-1}X \, a_{d\Delta} (\sigma Z)^{\Delta - d} \log \vert \sigma Z \vert \,
\phi_{0}^{\rm Poincare}\,.
\end{equation}
In appendix \ref{Poincare3} we reproduce this result for $d=2$, starting from a Poincar\'e mode sum.

\section{Rindler smearing in AdS${}_3$} \label{rads}

We'll work in AdS${}_3$ in Rindler coordinates, with metric
\[
ds^2 = - {r^2 - r_+^2 \over R^2} dt^2 + {R^2 \over r^2 - r_+^2} dr^2
+ r^2 d\phi^2\,.
\]
Here $-\infty < t,\phi < \infty$ and $r_+ < r < \infty$.  $R$ is the
AdS radius and $r_+$ is the radial position of the Rindler horizon.
With the ansatz $\phi(t,r,\phi) = e^{-i \omega t} e^{i k \phi}
f_{\omega k}(r)$ a normalizeable solution to the scalar wave equation
is \cite{Ichinose:1994rg,Keski-Vakkuri:1998nw}\footnote{The other
solution to the differential equation grows like $r^{\Delta - d}$.}
\[
f_{\omega k}(r) = r^{-\Delta}
                  \left({r^2 - r_+^2 \over r^2}\right)^{-i \hw / 2}
                  F\left({\Delta - i \hw - i \hk \over 2},
                    {\Delta - i \hw + i \hk \over 2},
                    \Delta,{r_+^2 \over r^2}\right)
\]
Here we define $\hw = \omega R^2 / r_+$ and $\hk = k R / r_+$.
Perhaps despite appearances, the mode functions $f_{\omega k}$ are
real and satisfy
\[
f_{\omega,k} = f_{\omega,-k} = f_{-\omega,k} = f_{-\omega,-k}\,.
\]
Note that $-\infty < \omega,k < \infty$ so, unlike global and
Poincar\'e, the Rindler modes involve a complete set of functions on
the boundary \cite{Unruh:1976db}.  This means we will have no freedom
in choosing the Rindler smearing function.  We therefore expect to
find boundary operators which are as well-localized as possible.

The field has an expansion in Rindler modes
\[
\phi(t,r,\phi) = \int_{-\infty}^\infty d\omega \int_{-\infty}^\infty dk \,
a_{\omega k} e^{-i \omega t} e^{i k \phi} f_{\omega k}(r)\,.
\]
The Rindler boundary field is given by
\[
\phi_0(t,\phi) = \lim_{r \rightarrow \infty} r^\Delta \phi(t,r,\phi) =
\int_{-\infty}^\infty d\omega \int_{-\infty}^\infty dk \, a_{\omega k}
e^{-i \omega t} e^{i k \phi}
\]
so we can express
\[
a_{\omega k} = {1 \over 4 \pi^2} \int dt d\phi \, e^{i \omega t} e^{-i k \phi} \phi_0(t,\phi)\,.
\]
We can therefore represent the bulk field in terms of the boundary field as
\be
\label{RindlerId}
\phi(t,r,\phi) = {1 \over 4 \pi^2} \int d\omega dk \, \left(\int dt' d\phi' \, e^{- i \omega (t - t')}
e^{i k (\phi - \phi')} \phi_0(t',\phi')\right) f_{\omega k}(r)\,.
\ee
If we were justified in changing the order of integration and doing
the integrals over $\omega$ and $k$ first, we would have an expression
for the Rindler smearing function which is just the Fourier transform
of the mode functions.
\begin{equation}
\label{DoesntWork}
K(t',\phi' \vert t,r,\phi) \mathop=^? {1 \over 4 \pi^2} \int d\omega dk \, e^{- i \omega (t - t')}
e^{i k (\phi - \phi')} f_{\omega k}(r)
\end{equation}
However the mode functions $f_{\omega k}$ diverge at large $k$, which
means we can't simply change the order of integration; we need to
proceed in a more careful way. We will find that a smearing function
can be constructed by analytically continuing to imaginary values of
the $\phi$ coordinate.

\subsection{Massless field in Rindler coordinates} \label{masslessrin}

In this subsection we specialize to a massless field ($\Delta = 2$),
and derive the smearing function in Rindler coordinates directly from
(\ref{RindlerId}) by a process of analytic continuation.  It will be
convenient to define the rescaled variables $\htt = r_+ t / R^2$, $\hp
= r_+ \phi / R$ and $\hw=wR^2/r_{+}$, $\hk=kR/r_{+}$. We'll also use
light-front coordinates
\[
\hw^\pm = {1 \over 2} (\hw \pm \hk) \, ,\qquad
\hx^\pm = \htt \pm \hp\,.
\]
Let us start by rewriting (\ref{RindlerId}) in the form
\bea
\label{RindlerId2}
\phi(t,r,\phi) & = & {1 \over 4 \pi^2} \int d\omega dk \, {1 \over \cosh (\pi \hw^+ / 2) \cosh (\pi \hw^-/2)} f_{\omega k}(r) \\
\nonumber
& & \left(\cosh (\pi \hw^+ / 2) \cosh (\pi \hw^-/2) \int dt' d\phi' \, e^{- i \omega (t - t')} e^{i k (\phi - \phi')}
\phi_0(t',\phi')\right)\,.
\eea
The point of breaking things up in this way is that it will lead to a
well-defined kernel, since $f_{\omega k} \lesssim e^{\pi k/2}$ as $k
\rightarrow \infty$.  We'll denote this modified kernel by $\tilde{K}$
below.

The hypergeometric function has an integral representation (\cite{Bateman}, p.~78))
\[
F(a,b,c,z) = {\Gamma(c) \over \Gamma(a) \Gamma(c - a)} {\Gamma(c) \over \Gamma(b) \Gamma(c - b)}
\int_0^1 ds \, \int_0^1 dt\, s^{a - 1} (1-s)^{c - a - 1} t^{b - 1} (1 - t)^{c - b - 1} (1 - stz)^{-c}\,.
\]
Specializing to a massless field ($\Delta = 2$) this becomes
\beas
F(1 - i \hat{\omega}^+, 1 - i \hat{\omega}^-, 2, r_+^2 / r^2) & = & {\sinh \pi \hw^+ \over \pi \hw^+}
{\sinh \pi \hw^- \over \pi \hw^-} \int_0^1 ds \, \int_0^1 dt \, \\
& & \left(s \over 1 - s \right)^{-i \hw^+} \left(t \over 1 - t \right)^{-i \hw^-} \left(1 - s t r_+^2 / r^2\right)^{-2}
\eeas
Using this to represent the mode functions in (\ref{RindlerId2}), the bulk field can be expressed as a convolution
\be
\label{RindlerConvolution}
\phi(\hx^+,\hx^-,r) = {1 \over r^\Delta} \int d\hy^+ d\hy^- \, \tilde{K}(\hx^+ - \hy^+,\hx^- - \hy^-,r) \,
\tilde{\phi}_0(\hy^+,\hy^-)
\ee
where
\bea
\nonumber
\tilde{K} & = & \int_0^1 ds \, \int_0^1 dt \, (1 - s t r_+^2 / r^2)^{-2} \int {d\hw^+ \over 2 \pi} {d\hw^- \over 2 \pi}
{\sinh (\pi \hw^+ / 2) \over \pi \hw^+ / 2} {\sinh (\pi \hw^- / 2) \over \pi \hw^- / 2} \\
& & \exp\left[-i \hw^+ (\hx^- - \hy^- + {1 \over 2} \log(1 - r_+^2/r^2) - \log {1 - s \over s}\right] \\
\nonumber
& & \exp\left[-i \hw^- (\hx^+ - \hy^+ + {1 \over 2} \log(1 - r_+^2/r^2) - \log {1 - t \over t}\right]
\eea
and where
\bea
\tilde{\phi}_0(\hy^+,\hy^-) & = & \int {d\hw^+ \over 2\pi} {d\hw^- \over 2\pi} \cosh (\pi \hw^+ / 2) \cosh (\pi \hw^-/2) \\
\nonumber
& & \left(\int d\hy'{}^+ d\hy'{}^- e^{-i \hw^+ (\hy^- - \hy'{}^-)} e^{-i \hw^- (\hy^+ - \hy'{}^+)} \phi_0(\hy'{}^+,\hy'{}^-) \right)\,.
\eea
The modified boundary field $\tilde{\phi}_0$ can be defined by analytic continuation, as we have
\beas
\tilde{\phi}_0(\hy^+,\hy^-) & = & \cosh \left({i \pi \over 2} {\partial \over \partial \hy^+}\right)
\cosh \left({i \pi \over 2} {\partial \over \partial \hy^-}\right) \phi_0(\hy^+,\hy^-) \\
& = & {1 \over 4} \Bigl( \phi_0(\hy^+ + {i \pi \over 2} - i \epsilon, \hy^- + {i \pi \over 2} - i \epsilon) +
\phi_0(\hy^+ + {i \pi \over 2} - i \epsilon, \hy^- - {i \pi \over 2} + i \epsilon) \\
& & \quad \phi_0(\hy^+ - {i \pi \over 2} + i \epsilon, \hy^- + {i \pi \over 2} - i \epsilon) +
\phi_0(\hy^+ - {i \pi \over 2} + i \epsilon, \hy^- - {i \pi \over 2} + i \epsilon) \Bigr)\,.
\eeas
This assumes the boundary field is analytic in the strip $-\pi/2 <
{\rm Im} \, \hy^+,\, {\rm Im} \, \hy^- < \pi/2$, which is true for
boundary fields constructed from finite superpositions of the global
boundary mode functions (\ref{gbmode}).

To compute the modified kernel $\tilde{K}$ it's convenient to first
act with $\partial_+ \partial_-$ to kill the $1/\hw^+ \hw^-$ factor,
and to define the $\sinh(\pi\hw^+/2) \sinh(\pi\hw^-/2)$ factor by
analytic continuation.  The $\hw^+$ and $\hw^-$ integrals then produce
$\delta$-functions which can be used to do the integrals over $s$ and
$t$.  Finally, upon integrating with respect to $\hx^+$ and $\hx^-$ we
find
\beas
\tilde{K}(\hx^+,\hx^-) & = & {4 \over \pi^2} {r^2 \over r_+^2} \sinh\left({i \pi \over 2} {\partial \over \partial \hx^+}\right)
            \sinh\left({i \pi \over 2} {\partial \over \partial \hx^-}\right) \\
& & \qquad  \log\left[\left(1 + \sqrt{1 - {r_+^2 \over r^2}} e^{\hx^+}\right)
                      \left(1 + \sqrt{1 - {r_+^2 \over r^2}} e^{\hx^-}\right)
                      - {r_+^2 \over r^2}\right] \\
& = & {r^2 \over \pi^2 r_+^2} \log { (1 - r_+^2 / r^2) \sinh^2 \htt + \cosh^2 \hp \over (1 - r_+^2 / r^2) \cosh^2 \htt + \sinh^2 \hp }
\eeas

Now let's go back to our expression for the bulk field
(\ref{RindlerConvolution}).  We'll break it up into two pieces, $\phi
= \int \tilde{K} \tilde{\phi}_0 = A + B$.  The first piece $A$
includes the terms in which the arguments of $\phi_0$ are shifted by
$\htt \rightarrow \htt \pm i \pi/2$.  That is
\beas
A & = & {1 \over 2 \pi^2 r_+^2} \int d\htt' d\hp' \, \log { (1 - r_+^2 / r^2) \sinh^2 (\htt - \htt') + \cosh^2 (\hp - \hp') \over
                                                    (1 - r_+^2 / r^2) \cosh^2 (\htt - \htt') + \sinh^2 (\hp - \hp')} \\
& & \qquad\qquad \left(\phi_0(\htt' + i \pi/2 - i\epsilon,\hp') + \phi_0(\htt' - i \pi/2 + i \epsilon,\hp')\right)
\eeas

\begin{figure}
\centerline{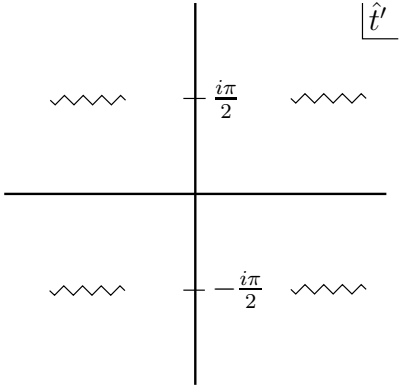}
\caption{Branch cuts in the $\htt'$ plane are located at ${\rm Im} \, \htt' = \pm \pi / 2$.}
\label{Tcuts}
\end{figure}

The logarithm has branch cuts indicated in Fig.~\ref{Tcuts}.  By
shifting the $\htt'$ contour of integration up or down by $i \pi/2$
one can make the arguments of $\phi_0$ real.  The imaginary parts of
the logarithm just above and below the cuts cancel, while the real
parts of the logarithm add to give
\be
\label{ShiftT}
A = {1 \over \pi^2 r_+^2} \int_{-\infty}^\infty d\htt' \int_{-\infty}^\infty d\hp'
\log \left\vert {- (1 - r_+^2 / r^2) \cosh^2 (\htt - \htt') + \cosh^2 (\hp - \hp') \over
                 - (1 - r_+^2 / r^2) \sinh^2 (\htt - \htt') + \sinh^2 (\hp - \hp')} \right\vert
\phi_0(\htt',\hp')
\ee

\begin{figure}
\centerline{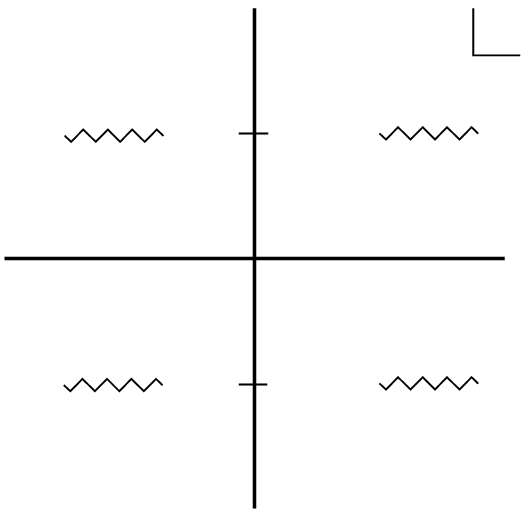 \hspace{2cm} 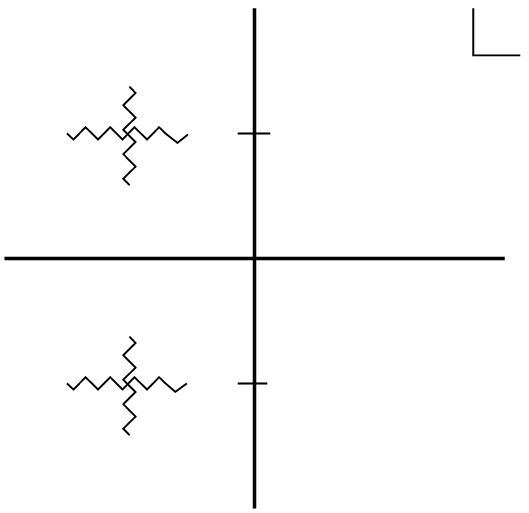}
\caption{For $\cosh (\htt - \htt') > 1/\sqrt{1 - r_+^2/r^2}$ the branch cuts in the $\hp'$ plane are at
${\rm Im} \, \hp' = \pm \pi / 2$ (left panel).  When $\cosh(\htt - \htt') = 1/\sqrt{1 - r_+^2/r^2}$ four of
the branch points touch, and for $\cosh(\htt - \htt') < 1/\sqrt{1 - r_+^2/r^2}$ the branch cuts are cross-shaped
(right panel).}
\label{Phicuts}
\end{figure}

Now consider the contribution $B$ coming from terms where $\hp \rightarrow \hp \pm i \pi / 2$, namely
\beas
B & = & {1 \over 2 \pi^2 r_+^2} \int d\htt' d\hp' \, \log { (1 - r_+^2 / r^2) \sinh^2 (\htt - \htt') + \cosh^2 (\hp - \hp') \over
                                                    (1 - r_+^2 / r^2) \cosh^2 (\htt - \htt') + \sinh^2 (\hp - \hp')} \\
& & \qquad\qquad \left(\phi_0(\htt',\hp' + i \pi / 2 - i \epsilon) + \phi_0(\htt',\hp' - i \pi/2 + i \epsilon)\right)
\eeas
The branch cuts of the logarithm are shown in Fig.~\ref{Phicuts}.  One
can push the $\hp'$ contour of integration up or down by $i \pi / 2$,
however if $\cosh (\htt - \htt') < 1/\sqrt{1 - r_+^2 / r^2}$ the
contour will get wrapped around the vertical part of the cut.  In
fact, after shifting $\hp'$, the contribution from horizontal part of
the $\hp'$ contour is
\be
\label{ShiftPhi}
B_{\rm horizontal} = {1 \over \pi^2 r_+^2} \int_{-\infty}^\infty d\htt' \int_{-\infty}^\infty d\hp' \log \left\vert
{- (1 - r_+^2 / r^2) \sinh^2 (\htt - \htt') + \sinh^2 (\hp - \hp') \over
 - (1 - r_+^2 / r^2) \cosh^2 (\htt - \htt') + \cosh^2 (\hp - \hp')} \right\vert
\phi_0(\htt',\hp')\,.
\ee
But this exactly cancels the contribution (\ref{ShiftT}) from shifting $\htt'$!  So the bulk field arises just
from the vertical part of the contour, that is from the discontinuity of the logarithm across the cut.  Setting
$\htt' = \htt + x$, $\hp' = \hp + i y$ this leads to
\beas
\phi(\htt,r,\hp) & = & {1 \over 2 \pi^2 r_+^2} \int_{-x_0}^{x_0} dx \biggl[
\int_{-\pi/2}^{-y_0} dy \, 2 \pi \phi_0(\htt + x,\hp + iy + i \pi/2) \\
& & \qquad\qquad\quad\,\,\, + \int_{y_0}^{\pi/2} dy \, 2 \pi \phi_0(\htt + x, \hp + i y - i \pi / 2)\biggr]
\eeas
where $\cosh x_0 = 1/\sqrt{1 - r_+^2 / r^2}$ and $\sin y_0 = \sqrt{1 - r_+^2 / r^2} \, \cosh x$.  Finally one can shift
$y \rightarrow y \pm \pi/2$ to obtain
\be
\label{MasslessSmear}
\phi(\htt,r,\hp) = {1 \over \pi r_+^2} \int_R dx dy \, \phi_0(\htt + x, \hp + i y)
\ee
where the region $R$ is defined by
\bea
\label{regionR}
&& \cosh x < 1/\sqrt{1 - r_+^2/r^2} \\
\nonumber
&& \cos y > \sqrt{1 - r_+^2/r^2} \, \cosh x\,.
\eea

It is worth emphasizing that we have been forced to work on the complexified
boundary.  Had there been a well-defined Rindler smearing function, with support on
the Rindler boundary at real values of the boundary coordinates, it would have been
given by (\ref{DoesntWork}).  Since that integral is divergent, no such smearing
function can exist.

\subsection{Wick rotating to de Sitter space} \label{cgeom}

In this subsection we derive the Rindler smearing function for general
conformal dimension.  Having seen that we need to analytically
continue the $\phi$ coordinate we begin by Wick rotating $\phi$ to
$\tilde \phi=i\phi$.  This gives a de Sitter geometry,
\[
ds^{2}=-\frac{r^{2}-r_{+}^{2}}{R^{2}}dt^{2}+\frac{R^{2}}{r^{2}-r_{+}^{2}}dr^{2}-r^{2}d\tilde{\phi}^{2}\,.
\]
Note that $-r$ plays the role of the time coordinate.  To avoid a
conical singularity at $r = 0$ we must periodically identify $\tilde
\phi\sim \tilde \phi +2 \pi R/r_+$.  The de Sitter invariant distance
function is
\[
\sigma=\frac{rr'}{r_{+}^{2}}\left(\cos\left(\frac{r_+(\tilde{\phi}-\tilde{\phi}')}{R}\right)-\sqrt{1-\frac{r_{+}^{2}}{r^{2}}}
\sqrt{1-\frac{r_{+}^{2}}{r'^{2}}}\cosh\left(\frac{r_+(t-t')}{R^2}\right)\right)~.
\]
We consider a scalar field of mass $m$ in de Sitter space.  For now we
take $m^2 R^2>1$, however later we will analytically continue $m^2 \to
- m^2$.  The analytically continued mass can be identified with the
mass of a field in AdS (note that the Wick rotation flips the
signature of the metric).

The field at some bulk point can be written in terms of the retarded
Greens function.  de Sitter space has numerous inequivalent vacuum
states, known as the alpha-vacua, which give rise to de Sitter
invariant correlators. The retarded Greens function is independent of
this choice of vacuum state.  It coincides with the imaginary part of
the commutator inside the past light-cone of the future point and
vanishes outside this region.  The field at some bulk point is
therefore
\be
\phi(r,\tilde{\phi},t)=\int d\tilde{\phi}'dt'\,\frac{r'(r'^{2}-r_{+}^{2})}{R^{2}}G_{ret}(r',\tilde{\phi}',t';r,\tilde{\phi},t)
\overleftrightarrow{\partial_{r'}}\phi(r',\tilde{\phi}',t') \label{brfield}
\ee
where the region of integration is over a spacelike surface of fixed
$r'$ inside the past light-cone of the bulk point.  In the $r' \rightarrow \infty$
limit this becomes the region $R$ introduced in (\ref{regionR}), namely
\bea
\label{plc}
&& \sqrt{1 - \frac{r_+^2}{r^2}} \cosh \frac{r_+( t-t')}{R^2} < 1 \\
\nonumber
&& \cos \frac{r_+( \tilde \phi-\tilde \phi')}{R} > \sqrt{1 - \frac{r_+^2}{r^2}} \, \cosh \frac{r_+( t-t')}{R^2}
\eea
As $r'\to\infty$ (with other coordinates held fixed) the retarded
Greens function takes the form \cite{Bousso:2001mw}
\[
G_{ret}\sim i\left( c\left(-\sigma -i\epsilon\right)^{-1+i
\sqrt{m^{2}R^2-1}}+c^* \left(-\sigma-i\epsilon\right)^{-1-i
\sqrt{m^{2}R^2-1}}-c.c.\right)
\]
where we take branch cuts along the positive real $\sigma$ axis and where
\[
 c=  \frac{\Gamma(2 i\sqrt{m^2-1}) \Gamma(1-i\sqrt{m^2-1}) }{2^{2-i
\sqrt{m^2-1}} R~ \Gamma(\frac{1}{2}+i\sqrt{m^2-1} ) }~. 
\]
The boundary field is defined as usual
\begin{equation}
\phi_{0}(\tilde{\phi},t)=\lim_{r\to\infty}r^{\Delta}\phi(r,\tilde{\phi},t)~.\label{eq:bfield}
\end{equation}
Choosing normalizable modes from the AdS viewpoint corresponds to
taking only positive frequencies in the $-r$ direction, which have a
$r^{-1-i\sqrt{m^2-1}}$ $r$-dependence.

Evaluating (\ref{brfield}) as $r' \to \infty$ we obtain the Rindler
smearing function\footnote{Here we use the identities $\sin\pi z =
\frac{\pi}{\Gamma(z)\Gamma(1-z)}$ and
$\frac{\Gamma(2z)}{\Gamma(z)\Gamma(1/2+z)}=\frac{2^{2z-1}}{\sqrt{\pi}}$.}
\bea
\nonumber
\phi(r,\phi,t) & = & {(\Delta - 1) 2^{\Delta - 2} \over \pi r_+^2} \int_R dx dy \,
\lim_{r' \rightarrow \infty} \left({\sigma \over r'}\right)^{\Delta - 2} \,
\phi_0(\phi + i{Ry \over r_+},t + {R^2 x \over r_+}) \\
& = & \frac{(\Delta-1)2^{\Delta-2}}{\pi r_{+}^{\Delta} }\int_{R}dx dy\left(\frac{r}{r_{+}}
\left(\cos y-\sqrt{1-\frac{r_{+}^{2}}{r^{2}}}\cosh x\right)\right)^{\Delta-2} \nonumber \\ 
&& \qquad\qquad \times \phi_{0}(\phi+i\frac{Ry}{r_+},t+\frac{R^2 x}{r_+})
\label{finalanswer}
\eea
In these expressions $\Delta=1+i\sqrt{m^2-1}$.  However by
analytically continuing $m^{2}\to-m^{2}$ we can take $\Delta$ to
coincide with the conformal dimension in AdS.  Since $\sigma>1$ in the
domain of integration this analytic continuation is straightforward.

As a check on this result, note that for $\Delta = 2$ we reproduce
(\ref{MasslessSmear}).  As a further check we can examine the limit $r
\to \infty$ where we should recover (\ref{eq:bfield}). In this limit
the region of integration becomes very small so we can Taylor expand
the smearing function, finding
\bea
\nonumber
\phi(r,\phi,t) & \sim & {(\Delta - 1) 2^{\Delta - 2} \over \pi r_+^\Delta} \phi_0(\phi,t)
\int_R dx dy \, {r^{\Delta - 2} \over r_+^{\Delta - 2}} \left({r_+^2 \over 2 r^2} - {1 \over 2} x^2 - {1 \over 2} y^2
\right)^{\Delta - 2} \\
& = & {1 \over r^\Delta} \phi_0(\phi,t)
\eea
as expected.

In this section we have used the fact that $\phi_0$ is analytic on a
strip in the complex $\phi$ plane centered on the real axis, which
will be true for fields built out of any superposition of a finite
number of global modes. The final result (\ref{finalanswer}) is
manifestly AdS covariant. We have checked that it is correct by
setting $\phi_0$ equal to a plane wave $e^{-i \omega t} e^{i k \phi}$
and numerically evaluating the integrals over $x$ and $y$, finding
values that agree with the corresponding bulk field $e^{-i \omega t}
e^{i k \phi} f_{\omega k}(r)$.

\section{Physical consequences} \label{physics}

\subsection{Bulk locality and UV/IR}

We have seen that one can define operators in the boundary theory that
in the large $N$ limit describe a free local bulk field. This is not
surprising, since this is the limit $l_{\rm Planck} \rightarrow 0$ where
classical supergravity is valid. The
construction of these operators in terms of a mode sum makes it
clear that by construction the two-point function of these CFT operators
will reproduce the bulk two-point function.
The two-point function is singular when bulk points are
coincident or are lightlike separated.  Since the smearing
functions are finite and have compact support, it's easy to see that
this singularity can only arise from UV singularities in the boundary
theory.\footnote{This was shown explicitly in \cite{hkll}.} This means that regions
inside the bulk are not related to a boundary theory with a conventional UV cutoff, so
there is no UV/IR relationship in the sense of relating bulk IR and
boundary UV cutoffs.  

What about scale-radius duality? In global and Poincar\'e coordinates
this duality is not manifest.  As can be seen in figure
\ref{GlobalFig} there may be a minimum smearing in the
time direction which is related to radial position in the bulk,
however the CFT operators are always completely smeared over the
spatial directions of the boundary.  In Rindler coordinates, on the other hand, we were
able to reduce the smearing integral to a compact region of the
complexified geometry, whose size shrinks to zero as the bulk point
approaches the boundary.  This makes scale-radius duality manifest.
For example, a bulk point at radius $r$ gets smeared over a range of
time $\delta t$ on the boundary given by
\begin{equation}
\cosh \frac{r_+ \delta t}{2R^2} = 1/\sqrt{1-r_{+}^{2}/r^{2}}\,.
\end{equation}
This is just the elapsed time between the point on the boundary which is
lightlike to the future of the bulk point at the same value of $\phi$,
and the point on the boundary which is lightlike to the past at the
same $\phi$. The smearing is also over some finite region in imaginary
$\phi$, as per (\ref{plc}). 

\subsection{Finite $N$ and the holographic bound}

We begin with a few remarks on the boundary commutator in the complexified
geometry of section \ref{cgeom}.  The boundary Wightman two-point
function is
\begin{equation}
\label{VacWight}
\langle \phi_0(t,\phi)\phi_0(\phi',t') \rangle \sim \left(\cosh \frac{r_+ \delta \phi}{R}-
\cosh\left(\frac{r_+\delta t}{R^2} - i\epsilon \right)\right)^{-2\Delta}\,.
\end{equation}
In the large $N$ limit, with free bulk fields,
the commutator is a c-number which vanishes whenever the $i\epsilon$ term can be
neglected.  We are interested
in real $t$ and complex $\phi$.  So at large $N$ the commutator is non-zero
in only two situations:
\begin{enumerate}
\item $\delta \phi$ real and $\delta t > R \delta \phi$ (the usual case of timelike separation),
\item $\delta \phi$ purely imaginary and $\delta t$ arbitrary,
\end{enumerate}
while for generic complex $\delta \phi$ the commutator vanishes. What
happens at finite $N$?  Since the vacuum two-point function is
determined by conformal invariance, (\ref{VacWight}) is true even at
finite $N$.  However the commutator is an operator rather than a
c-number, so we cannot conclude that the commutator (rather than its
vacuum expectation value) vanishes.  Still, it seems reasonable to
assume that up to $1/N$ corrections to the size of the regions, the
commutator will be non-zero only if condition 1 or 2 is satisfied.
  
At infinite $N$ the smeared operators we have constructed commute when
the bulk points are spacelike separated.  This works, even when the
smeared operators overlap on the boundary, because the commutator of
the boundary operators is a c-number rather than an operator.  At
finite $N$ this picture must change in an interesting way if a
holographic description is to be maintained.  Commutators of boundary
operators become operators rather than c-numbers, which destroys the
delicate balance that enabled two operators smeared over regions
timelike to each other to commute. While the generalization of bulk
field operators to finite $N$ is very difficult \cite{Banks:1998dd},
we nevertheless get tight constraints from the $N\rightarrow \infty$
limit.

We argue that the only generic way for two smeared operators to
commute at finite $N$ is by smearing over disjoint ``spacelike''
(commuting) regions on the boundary.  This motivates representing
local bulk operators using a form of the smearing function with
minimal spread on the boundary, in the hope that such boundary
operators provide the ``most local'' definition of bulk operators at
finite $N$.  We proceeded to this goal in a series of steps, first
reducing from smearing operators over the entire boundary, to only
smearing over points spacelike separated from the bulk point, and
finally to smearing over a compact region of the complexified
geometry.\footnote{Although we explicitly performed this last step
only for three dimensions in Rindler coordinates, it is
straightforward to generalize to arbitrary dimensions and other
coordinates systems.}  This prepares us to count the number of
independent commuting bulk operators inside a given volume.

Consider two local bulk operators at the same values of $r$ and $t$
but different $\phi$. Up to $1/N$ corrections to the actual size of
the region, these will correspond to boundary operators smeared in the
$t$ and imaginary $\phi$ directions according to (\ref{plc}). It
therefore is reasonable to assume that even at finite $N$ these
operators will commute if the $\phi$ separation is sufficiently large
that the boundary commutator always vanishes. This requires
\begin{equation}
\label{dphisep}
\cosh \frac{r_+\delta \phi}{2 R} > 1 / \sqrt{1-r^{2}_{+}/r^{2}}\,.
\end{equation}
Let us work at large $r$.  Then we expect bulk operators separated by
$\delta \phi = 2R/r$ to commute at finite $N$.  Consider the set of
such operators at fixed $r$ and $t$. Operators at smaller values of
$r$ and the same $t$ will be smeared over a larger time interval on
the boundary, so will not trivially commute with this set. Then the
number of trivially commuting operators that can be localized to a
radius $\leq r$, per radian along the boundary, per independent CFT
degree of freedom is $r/2R$.  Heuristically the number of CFT degrees
of freedom is given by the central charge, so the maximum number of
commuting operators per radian is of order
\begin{equation}
cr/2R\,.
\end{equation}
This result is consistent with expectations from holography, and gives
a nice picture of how the number of commuting degrees of freedom is
drastically reduced. This also makes it clear where canonical
quantization of gravity fails: the degrees of freedom on a Cauchy
hypersurface do not commute.  Note that if $\phi$ is periodically
identified to give a BTZ black hole, $\phi \sim \phi + 2 \pi$, then
this counting breaks down when $2R/r \approx 2\pi$ and the operator is
smeared over the entire boundary.  Such a breakdown is expected, since
a Hawking-Page transition occurs for a black hole of radius $r \sim R$
\cite{Birmingham:2002ph}.

\bigskip
\centerline{{\bf Acknowledgements}}
\noindent
We thank Lenny Susskind and Shubho Roy for valuable discussions.  The
work of AH and DK is supported by US Department of Energy grant
DE-FG02-92ER40699.  GL is supported in part by a grant from the
Israeli science foundation number 568/05.  The research of DL is
supported in part by DOE grant DE-FG02-91ER40688-Task A.

\appendix
\section{Even AdS: global Greens function} \label{eggads}

In this section we show how to reproduce our global smearing function
in even-dimensional AdS starting from a spacelike Greens function.  To
construct such a Greens function we first find the general (singular)
AdS-invariant solution to the homogeneous wave equation in Euclidean
space.  The solution involves two arbitrary constants.  We fix one
constant by requiring that the solution is in fact a Greens function
with a properly-normalized delta-function source at the origin.  We
fix the other constant by requiring that, upon analytically continuing
to Lorentzian AdS, the Greens function is non-zero only at spacelike
separation.

The AdS-invariant distance (\ref{DistanceFtn}) is defined for
Lorentzian AdS.  However by Wick rotating $\tau = - i \tau_E$ one can
also use $\sigma$ as an invariant distance function on Euclidean AdS;
continuing back to Lorentzian signature corresponds to the
prescription $\sigma \rightarrow \sigma + i \epsilon$.  For
AdS-invariant fields\footnote{Where we use the notation $\phi(x) =
\phi(\sigma(x \vert x'))$.} the wave equation $(\Box - m^2)\phi = 0$
reduces to
\be
\label{WaveEqn}
(\sigma^2 - 1) \phi'' + (d+1) \sigma \phi' - \Delta (\Delta - d) \phi = 0\,.
\ee
The general solution is
\be
\phi(\sigma) = c_1 (\sigma^2 - 1)^{-\mu/2} P_\nu^\mu(\sigma) + c_2 (\sigma^2 - 1)^{-\mu/2} Q_\nu^\mu(\sigma)
\ee
where $P^\mu_\nu$, $Q^\mu_\nu$ are associated Legendre functions with
$\mu = {D - 2 \over 2}$, $\nu = \Delta - {D \over 2}$.  In even AdS
note that $\mu$ is a non-negative integer, in which case as $\sigma
\rightarrow 1$ the Legendre functions have the asymptotic behavior
\cite{Bateman}
\begin{equation}
P^\mu_\nu(\sigma) \sim {2^{-\mu/2} \Gamma(\nu + \mu + 1) (\sigma - 1)^{\mu/2} \over \mu! \, \Gamma(\nu - \mu + 1)}\, , \quad
Q^\mu_\nu(\sigma) \sim 2^{\mu/2 - 1} \Gamma(\mu)  e^{i \pi \mu} (\sigma - 1)^{-\mu/2}\,.
\end{equation}
A Euclidean Greens function should have the short-distance behavior
\begin{equation}
\label{EuclideanShort}
G_E(r) \sim - {1 \over (D - 2) {\rm vol}(S^d) r^{D-2}} \quad \hbox{\rm as $r \rightarrow 0$}
\end{equation}
where $r$ is a Euclidean radial coordinate and ${\rm vol}(S^d) = 2
\pi^{D/2}/\Gamma(D/2)$.  At short distances $\sigma \approx 1 + r^2/2
R^2$.  So $\phi(\sigma)$ will be a Euclidean Greens function with a
properly normalized (unit-strength) delta-function source at the
origin provided
\be
c_1 = {\rm arbitrary}\, , \qquad c_2 = {(-1)^{\mu + 1} \over 2^{\mu - 1} (D-2) {\rm vol}(S^d) \Gamma(\mu) R^{D-2}}\,.
\ee
Wick rotating back to Lorentzian AdS we set $G_M(\sigma) = i \phi(\sigma + i \epsilon)$ so that
\begin{equation}
\label{MinkowskiGreens}
(\Box - m^2) G_M = {1 \over \sqrt{-g}} \delta^D(x)\,.
\end{equation}
$G_M$ has the same short-distance behavior as the standard Feynman
Greens function, although we have not yet fixed its large-distance
behavior (which depends on $c_1$).

We choose $c_1$ to make the Greens function vanish at timelike
separation.  With a $\sigma \rightarrow \sigma + i \epsilon$
prescription the analytic continuation into the so-called ``cut''
region $-1 < \sigma < 1$ is \cite{Bateman}
\begin{equation}
G_M(\sigma) = i c_1 (-1)^\mu (1 - \sigma^2)^{-\mu/2} \hat{P}^\mu_\nu(\sigma) + i c_2 (-1)^\mu (1 - \sigma^2)^{-\mu/2}
\left(\hat{Q}^\mu_\nu(\sigma) - {i \pi \over 2} \hat{P}^\mu_\nu(\sigma)\right)
\end{equation}
where $\hat{P}$, $\hat{Q}$ are variants of the associated Legendre
functions (denoted with upright $P$'s and $Q$'s in \cite{Bateman})
which are real for $-1 < \sigma < 1$.  Since (\ref{MinkowskiGreens})
is a real equation we only need to keep the real part of $G_M$.  On the interval
$-1 < \sigma < 1$ this is given by
\begin{equation}
{\rm Re} \, G_M(\sigma) = \left({\rm Re}(i c_1) + {\pi c_2 \over 2}\right)
(-1)^\mu (1 - \sigma^2)^{-\mu/2} \hat{P}^\mu_\nu(\sigma)\,.
\end{equation}
Note that ${\rm Re} \, G_M$ vanishes for $-1 < \sigma < 1$ provided $c_1 = i \pi c_2 / 2$.  With this choice we
can construct a new Greens function $G$ which is non-zero only
at spacelike separation:\footnote{$G_M(x \vert x')$ also has delta
function sources at all the $2 \pi$ images of the point $x'$.  By
restricting $G$ to points that are spacelike separated from $x'$ we keep
only a single source.}
\bea
\nonumber
G(x \vert x') & \equiv & \left\lbrace
\begin{array}{ll}
{\rm Re} \, G_M(x \vert x')  &   \quad \hbox{\rm at spacelike separation} \\
0  &   \quad \hbox{\rm otherwise}
\end{array}
\right. \\
& = & - {\pi c_2 \over 2} (\sigma^2 - 1)^{-\mu/2} P^\mu_\nu(\sigma) \theta({\rm spacelike}) \,.
\eea
We can plug this into Green's identity
\begin{equation}
\phi(x') = \int d\tau d\Omega \sqrt{g_\Omega} {R^{D-2} \over \cos^{D-2} \rho} \left(\phi \partial_\rho G - G \partial_\rho \phi\right)
\vert_{\rho \rightarrow \pi/2}
\end{equation}
to obtain the corresponding smearing function.  Noting the asymptotic behavior
\begin{equation}
P^\mu_\nu(\sigma) \sim {2^\nu \Gamma(\nu + 1/2) \sigma^\nu \over \sqrt{\pi} \Gamma(\nu-\mu+1)} \quad
\hbox{\rm as $\sigma \rightarrow \infty$}
\end{equation}
we have
\begin{equation}
\phi(x') = \int d\tau d\Omega \sqrt{g_\Omega} \, K(x \vert x') \phi_0(x)
\end{equation}
where the smearing function is
\begin{equation}
\label{GlobalEvenGreens}
K(x \vert x') = {(-1)^{(D-2)/2} 2^{\Delta - D} \Gamma(\Delta - {d \over 2} + 1) \over \pi^{d/2} \Gamma(\Delta - d + 1)}
\lim_{\rho \rightarrow \pi/2} \left(\sigma(x \vert x') \cos \rho \right)^{\Delta - d} \theta({\rm spacelike})\,.
\end{equation}
This agrees with the result (\ref{GlobalEvenSmear}) obtained from a global mode sum in even AdS.

\section{AdS covariance in odd dimensions} \label{covariance}

To show that the smearing function is AdS covariant in odd dimensions
we must show that
\begin{equation}
\label{CovarianceIdentity}
\int d\tau d \Omega \, \lim_{\rho_0 \rightarrow \pi/2}
(\sigma \cos \rho_0)^{\Delta - d} \ln J \, \phi_0^{global} = 0
\end{equation}
where $J$ is the Jacobian on the boundary induced by an AdS
transformation of the bulk.  Consider an AdS isometry which takes a
point in the bulk to $\rho = 0$.  To compute the corresponding
Jacobian introduce the embedding coordinates
\begin{eqnarray}
Y_0 &=& R \sec \rho \cos \tau = \frac{1}{2Z} (R^2 + r^2 + Z^2 - T^2) \label{embed_1}\\
Y_1 &=& R \sec \rho \sin \tau = R \frac{T}{Z} \\
X_0 &=& R \tan \rho\  w_0 = \frac{1}{2Z} (R^2 - r^2 - Z^2 + T^2) \\
\vec{X} &=& R \tan \rho \ \vec{w} = R \frac{\vec{X}}{Z} \label{embed_4}
\end{eqnarray}
where $w_0, \vec{w}$ are coordinates on a $(D-2)$--sphere.  AdS
transformations are given by rotations and boosts in these
coordinates.  The center of AdS, at $\rho=0$, is given in the
embedding coordinates by $X_0=\vec{X}=0$.  Given an arbitrary point in
the bulk we can set $\vec{X} = Y_1 = 0$ by performing angular
rotations and $\tau$ translations in global coordinates, for which the
Jacobian is unity.  We then boost in the $X_0-Y_0$ plane, resulting in a
change of coordinates
\begin{eqnarray*}
X_0' &=& -\sinh \alpha Y_0 + \cosh \alpha X_0 \\
Y_0' &=& \cosh \alpha Y_0 - \sinh \alpha X_0 \;.
\end{eqnarray*}
Since we want $X_0'=0$, this determines the parameter $\tanh \alpha =
X_0/Y_0$.  Now, turning our attention to the boundary, the Jacobian of
the transformation is
\begin{eqnarray}
\nonumber
J & = &\lim_{\rho_0 \to \pi/2} \cos \rho_0 / \cos \rho_0' \\
\Rightarrow \quad J^2 & = & \frac{Y_0^{'2} + Y_1^{'2}}{Y_0^2 + Y_1^2} \\
\nonumber
& = & \frac{(\cosh \alpha Y_0 - \sinh \alpha X_0)^2 + Y_1^2}{Y_0^2 + Y_1^2}\,.
\end{eqnarray}

The strategy is to show that the integrand of
(\ref{CovarianceIdentity}) is analytic in the lower half complex
plane, making the contour integral vanish.  We find it convenient to
perform the calculation in a variation on lightfront Poincar\'e
coordinates, given by $X^\pm = T \pm r$ with $r=|\vec{X}|$.  However
we have to be careful regarding the domain of integration of $X^\pm$,
since the spatial distance $r \ge 0$.  If we restrict the domain of
the angular coordinates on the boundary to cover only half the $d-2$
sphere, and instead allow $-\infty < r < \infty$, then $X^\pm$ has the
full range of integration.  One note, however.  The integration
measure in these new coordinates is proportional to $|r|^{d-2}$.  In
odd dimensional AdS, where $d$ is even, the measure factor is
analytic.  This would not be true in even dimensional AdS.

As in section \ref{sect:OddAdSPoincare} we project all boundary points into one
Poincar\'e patch.  Referring to (\ref{embed_1}) - (\ref{embed_4}) we
want to evaluate the integral
\begin{eqnarray*}
I &=& \int d^dx (\sigma Z_0)^{\Delta - d} \left. \right|_{T \to T - i\epsilon} \ln \left( \frac{\cos \rho_0'}{\cos \rho_0} \right) \phi_{0+}^{Poincare} + \mathrm{c.c.} \\
&=& \frac{1}{2} \int d^dx (\sigma Z_0)^{\Delta - d} \left. \right|_{T \to T - i\epsilon}\left\{ \ln \left[ (R^2 + r^2 - T^2)^2 + 4 R^2 T^2 \right] \right. \\
&-& \left. \ln \left[ \left( \cosh \alpha(R^2 + r^2 - T^2) - \sinh \alpha (R^2 - r^2 + T^2) \right)^2 + 4 R^2 T^2 \right] \right\} \phi_{0+}^{Poincare} + \mathrm{c.c.} \\
&=&\frac{1}{2} \int d^dx (\sigma Z_0)^{\Delta - d} \left. \right|_{T \to T - i\epsilon}\left\{ \ln \left(R^2 + (X^+)^2\right) + \ln \left( R^2 + (X^-)^2 \right) \right. \\
&-&\left. \ln \left( e^\alpha R^2 + e^{-\alpha} (X^+)^2 \right) - \ln \left( e^\alpha R^2 + e^{-\alpha} (X^-)^2 \right) \right\} \phi_{0+}^{Poincare} + \mathrm{c.c.}
\end{eqnarray*}

We now show that the integrand is analytic in the lower half plane of
one of the lightfront coordinates.  With our $T \to T - i\epsilon$
prescription the branch points of $(\sigma Z_0)^{\Delta - d}$ are in
the upper half complex plane of both $X^\pm$.  Each log term is
independent of either $X^+$ or $X^-$, and so is trivially analytic in
that coordinate.  Finally the boundary field contains terms like
$e^{-i(\omega^+ X^- + \omega^- X^+)}$ where $\omega^\pm = \omega \pm
|k| \cos \theta$, $\theta$ being the angle between $\vec{X}$ and the
momentum $\vec{k}$.  Note that $\omega^\pm \ge 0$, due to the fact
that $\omega \ge |k|$, so the boundary field is analytic in the lower
half complex plane of both $X^\pm$.  This shows that the contour
integral over one of the lightfront coordinates is zero, so $I$
vanishes and the smearing function is AdS covariant.

The same procedure can be used when converting the smearing function
from global coordinates to Poincar\'e, showing that
\[
\int d^dx \, (\sigma Z_0)^{\Delta - d} \ln \left( \frac{Z_0}{\cos \rho_0} \right) \phi_0^{Poincare} = 0 \;.
\]

\section{Odd AdS: Poincar\'e mode sum} \label{Poincare3}

In Poincar\'e coordinates it is possible to construct a smearing
function by directly evaluating the Poincar\'e mode sum.  Bena did
this in AdS${}_5$ \cite{Bena}; here we'll do the analogous calculation
in AdS${}_3$.

In Poincar\'e coordinates the mode expansion of a real scalar field is
\be
\label{PoincareModeExp}
\phi(T,X,Z) = \int_{\omega > \vert k \vert} d\omega dk \, a_{\omega k} e^{-i \omega T} e^{i k X}
Z J_\nu(\sqrt{\omega^2 - k^2} Z) + {\rm c.c.}
\ee
where $J_\nu$ is a Bessel function of order $\nu = \Delta - 1$.  The Poincar\'e boundary field is
\beas
\phi_0(T,X) & = & \lim_{Z \rightarrow 0} {1 \over Z^\Delta} \phi(T,X,Z) \\
& = & {1 \over 2^{\Delta - 1} \Gamma(\Delta)} \int_{\omega > \vert k \vert} d\omega dk \,
a_{\omega k} \left(\omega^2 - k^2\right)^{(\Delta - 1)/2} e^{-i \omega T} e^{i k X}\,.
\eeas
Thus we can express the bulk field in terms of the boundary field,
\be
\phi(T,X,Z) = \int dT' dX' \, K(T',X' \vert T,X,Z) \phi_0(T',X')
\ee
where the smearing function is
\beas
K(T',X' \vert T,X,Z) & = & {2^{\Delta - 3} \Gamma(\Delta) Z \over \pi^2}
\int_{\omega > \vert k \vert} d\omega dk \,
e^{- i \omega (T - T')} e^{i k (X - X')} \\
& & \qquad {1 \over (\omega^2 - k^2)^{(\Delta - 1)/2}} \,
J_\nu(\sqrt{\omega^2 - k^2} \, Z) + {\rm c.c.}
\eeas
To keep the integral convergent we should give $T'$ a positive imaginary part.

It's straightforward to evaluate the positive-frequency part of the smearing function.
It suffices to set $T = X = X' = 0$ and consider
\be
K_+(T',0 \vert 0,0,Z) = {2^{\Delta - 3} \Gamma(\Delta) Z \over \pi^2}
\int_{\omega > \vert k \vert} d\omega dk \,
e^{i \omega T'} {1 \over (\omega^2 - k^2)^{(\Delta - 1)/2}} \,
J_\nu(\sqrt{\omega^2 - k^2} Z)
\ee
Setting
\be
\omega^+ = {1 \over 2} (\omega + k) = r e^{\xi} \qquad\quad
\omega^- = {1 \over 2} (\omega - k) = r e^{-\xi}
\ee
we have
\beas
K_+ & = & {2^{\Delta - 1} \Gamma(\Delta) Z \over \pi^2} \int_0^\infty r dr \,
\int_{-\infty}^\infty d\xi \, e^{i 2 r T' \cosh \xi} {1 \over (2 r)^\nu} J_\nu(2 r Z) \\
& = & {2^{\Delta - 2} \Gamma(\Delta) Z \over \pi^2} \int_0^\infty dr \,
{1 \over r^{\nu - 1}} K_0(-i r T') J_\nu(rZ) \\
& = & - {Z^\Delta \over 2 \pi^2 T'{}^2} F(1,1,\Delta,Z^2/T'{}^2)\,.
\eeas
The Lorentz-invariant generalization is
\be
\label{PoincareModeSum}
K_+(T',X' \vert 0,0,Z) = - {1 \over 2 \pi^2} \, {Z^\Delta \over T'{}^2 - X'{}^2}
F \left(1,1; \Delta; {Z^2 \over T'{}^2 - X'{}^2}\right)
\ee
where again the singularities are to be handled with a $T' \rightarrow
T' + i \epsilon$ prescription.  Since $K_+$ is constructed from
positive-frequency modes its complex conjugate $K_-$ only involves
negative frequency modes.  Then $\int K_- \phi_{0+}$ vanishes, and we
can take the full smearing function to be given by $K = K_+ + K_-$.

Note that the smearing function we have constructed has support on the
entire boundary of the Poincar\'e patch.  Also it can be applied
directly to the boundary field $\phi_0$; unlike the smearing function
constructed in section \ref{sect:OddAdSPoincare} one does not have to
decompose $\phi_0$ into its positive and negative frequency
components.  It does have one drawback, however: the smearing function
we have constructed is not AdS-covariant.

One might ask how the Poincar\'e mode sum is related to the covariant
results obtained in section \ref{sect:OddAdSPoincare}.  This is
easiest to understand when $\Delta$ is an integer, in which case one
has
\begin{eqnarray}
& & \xi F(1,1,\Delta,\xi) = - (\Delta - 1) \left(1 - {1 \over \xi}\right)^{\Delta - 2} \log(1 - \xi) \\
\nonumber
& & \qquad\qquad\qquad + (\hbox{\rm polynomial of degree $\Delta - 3$ in $1/\xi$})
\end{eqnarray}
Applying this to (\ref{PoincareModeSum}) gives
\bea
\nonumber
K_+ & = & {\Delta - 1 \over 2 \pi^2} \left({-T'^2 + X'^2 + Z^2 \over Z}\right)^{\Delta - 2}
\log {T'^2 - X'^2 - Z^2 \over T'^2 - X'^2} \\
& & \qquad + \big(\hbox{\rm polynomial in ${T'^2 - X'^2 \over Z^2}$}\big)
\eea
We can drop the polynomial, since it vanishes when integrated against $\phi_{0+}$ (close the contour in the
lower half $T'$ plane).  Also we can write
\bea
\nonumber
K_+ & = & {\Delta - 1 \over 2 \pi^2} \left({-T'^2 + X'^2 + Z^2 \over Z}\right)^{\Delta - 2}
\log {T'^2 - X'^2 - Z^2 \over 2 Z} \\
& & \qquad + ({\rm polynomial}) \cdot \log {(T'+X')(T'-X') \over 2Z}
\eea
The second line vanishes when integrated against $\phi_{0+}$.  To see this recall that the Poincar\'e mode
expansion (\ref{PoincareModeExp}) only involves modes with $\omega > \vert k \vert$, and close the integration
contour in the lower half of the $T'+X'$ or $T'-X'$ plane as appropriate.  Then we are left with
\bea
K_+ & = & {\Delta - 1 \over 2 \pi^2} \left({-T'^2 + X'^2 + Z^2 \over Z}\right)^{\Delta - 2}
\log {T'^2 - X'^2 - Z^2 \over 2 Z} \\
& = & {(\Delta - 1) 2^{\Delta - 2} \over 2 \pi^2} \lim_{Z' \rightarrow 0} (\sigma Z')^{\Delta - 2} \log (\sigma Z')\,.
\eea
We can replace $K_+ \rightarrow K_+ + {\rm c.c.}$, since the complex conjugate drops out when integrated
against $\phi_{0+}$.  This leaves
\begin{equation}
K = K_+ = K_- = {(\Delta - 1) 2^{\Delta - 2} \over \pi^2} \lim_{Z' \rightarrow 0} (\sigma Z')^{\Delta - 2}
\log \vert \sigma Z' \vert
\end{equation}
in agreement with (\ref{IntegerDeltaPoincare}) for $d = 2$.

\providecommand{\href}[2]{#2}
\begingroup\raggedright

\endgroup

\end{document}